\documentclass[11pt,a4paper]{article}
\begin{document} 
\baselineskip 5.4mm
\begin{center}
{\Large \bf
Supersymmetric Theory of (Color)superconductivity
}
\vspace{2mm}\\
Tadafumi Ohsaku
\end{center}

A few years ago, I published papers on the examinations of the theory of BCS superconductivity
in systems ( metal, metalic hydrogen, plasma, neutron star, etc. ) 
they have various numerical values of chemical potential $\mu$~[1,2,3].
It was shown that the BCS behavior is quite universal under the variation of chemical potential.  
In this talk, I discuss the extension of these previous publications to the case 
where the effect of SUSY, namely superpartner, might appear clearly.
We consider (color)superconductivity ( (C)SC ) of a condensed matter with 
$\mu\sim{\cal O}(\Delta)$, where $\Delta$ is the SUSY breaking scale ( $\sim$ TeV ).
Recently, attempts to examine the theories of 
CSC from the viewpoint of SUSY appeared in literature~[4,5].
According as the results of the nonperturbative method of SUSY gauge theories, 
Ref.~[4] discussed the symmetry breaking patterns of the ${\cal N}=1$ SQCD 
at $\mu\ne 0$, and $\Delta\ne 0$ for squarks.
Because the validity of the exact results of SUSY gauge theories to the problem of CSC is not clear, 
the method of Ref.~[4] could apply to the situation of $\mu < \Delta \ll \Lambda_{SQCD}$, 
where $\mu$ and $\Delta$ could be regarded as a small perturbation to a SUSY gauge theory.
The purpose and/or situation of this work are different from that of Refs.~[4,5]. 

The summary of the method of this work: 
(1) The case of both $U(1)$ and $SU(N_{c})$ gauge symmetries are examined.
(2) Both the ${\cal N}=1$ 4D and ${\cal N}=2$ 3D cases are considered.
(3) To construct a BCS-type theory, we employ the ${\cal N}=1$ 4D 
generalized SUSY Nambu$-$Jona-Lasinio model with $\mu\ne0$ as the model Lagrangian.
(4) The ${\cal N}=2$ 3D theory is obtained by a dimensional reduction 
of the 4D counterpart. 
(5) After adopting SUSY auxiliary fields for diquark-type composites,
the effective potential and the gap equation are obtained by the method of large-$N$ expansion.

The summary of the results: 
(1) By examining the quasiparticle excitation energy spectra and the gap equations, 
the conditions for Bose-Einstein condensation of the scalar sector, dynamical chiral symmetry breaking (DCSB),
and (color)superconductivity were found.
(2) The effects of the bosonic part in the gap equation in the SUSY BCS theory have been reviewed,
while there is no SUSY effect in the thermodynamic character of the SUSY (C)SC.
(3) The (C)SC shows the BCS character even if $\mu$ is close to $\Delta$.
(4) There is no critical coupling also in the SUSY BCS case, 
while we have found the critical soft mass $\Delta_{cr}$:
SUSY protects not only the chiral invariance from DCSB but also
the dynamical breaking of gauge symmetry at $\mu\ne 0$.
(5) The BCS universal constant $gap/T_{c}=1.76$
is always satisfied both in the ${\cal N}=1$ 4D and the ${\cal N}=2$ 3D cases.
The results were published in Refs.~[6,7].

\end{document}